\documentclass[a4paper,12pt]{article}
\usepackage{amssymb,amsmath}
\usepackage{epsfig,fancyheadings}

\usepackage{a4}
\usepackage{parskip}
\usepackage{color,soul,rotating}
\usepackage{amsthm,amsxtra,amscd,relsize,multirow}
\usepackage[all,2cell]{xy}\UseAllTwocells

\newcommand{\sect}[1]{\setcounter{equation}{0}\section{#1}}
\renewcommand{\theequation}{\arabic{section}.\arabic{equation}}

\def\N{{\mathcal N}}
\def\L{{\mathcal L}}

\def\L{{\mathcal L}}

\def\H{{\mathcal H}}

\def\a{\alpha}

\def\d{\delta}

\def\s{\sigma}

\def\m{\mu}
\def\n{\nu}

\def\f{\phi}
\def\vf{\varphi}

\def\l{\lambda}

\def\o{\omega}
\def\sinh{\mathrm{sinh}}
\def\cosh{\mathrm{cosh}}

\def\p{\partial}
\def\rb{\right}
\def\lb{\left}
\def\axs{AdS_5\times S^5}
\newcommand{\eq}[1]{\begin{equation} #1 \end{equation}}
\newcommand{\al}[1]{\begin{align} #1 \end{align}}
\newcommand{\ml}[1]{\begin{multline} #1 \end{multline}}

\def\cp{\mathbb {CP}^3}

\begin{titlepage}
\title{Near Flat Space limit of strings on $AdS_4\times\cp$}

\author{M.~Kreuzer\thanks{e-mail:kreuzer@hep.itp.tuwien.ac.at}, 
R.~C.~Rashkov\thanks{e-mail:rash@hep.itp.tuwien.ac.at; 
on leave from Dept of Physics, Sofia University, Bulgaria.} 
and M.~Schimpf
\ \\ \ \\
 Institute for Theoretical Physics, \\ Vienna
University of Technology,\\
Wiedner Hauptstr. 8-10, 1040 Vienna, Austria
}
\date{}
 \end{titlepage}

\begin{document}

\maketitle
\thispagestyle{fancy}

 \begin{abstract}
The non-linear nature of string theory on non-trivial backgrounds related to the AdS/CFT correspondence 
suggests to look for simplifications. Two such simplifications proved to be useful in studying string theory. These are 
the pp-wave limit which describes point-like strings and the so called ``near flat space'' limit which connects 
two different sectors of string theory -- pp-waves and ``giant magnons''. 
Recently another example of AdS/CFT duality emerged - $AdS_4/CFT_3$,
which suggests duality between $\mathcal N=6$ CS theory and superstring theory on $AdS_4\times \cp$.
In this paper we study the ``near flat space'' limit of strings on the $AdS_4\times \cp$ background and discuss possible
applications of the reduced theory.
 \end{abstract}

\sect{Introduction}

The idea of a correspondence between the large N limit of gauge
theories and strings emerged over thirty years ago
\cite{'tHooft:1973jz} and since then many attempts for the explicit realization
have been made. One of the most promissing so far was
provided by Maldacena who conjectured the AdS/CFT correspondence
\cite{holography}. Since then this became a topical area
and many fascinating discoveries were made over the last decade.
One of the predictions of the correspondence
is a one-to-one map between the spectrum of  string theory on
$\axs$ and the spectrum of anomalous dimensions of gauge invariant
operators in $\N=4$ Supersymmetric Yang-Mills (SYM)
theory.

The vast majority of papers were on qualitative and quantitative
description of $\mathcal{N}=4$ SYM theory with $SU(N)$ gauge group
by making use of the string sigma model on $AdS_{5}\times S^{5}$. This special case
is remarkable in several ways. First of all the background preserves
the maximal number of supersymmetries. Secondly, as indicated first in \cite{BPR},
the huge amount of symmetries of the relevant classical coset sigma model leads
to its integrability and opens the possibility to study its integrability
at the quantum level as well.
Thus, one can put forward the important issue of derivation of the S-matrix and 
study the main features of the string model and its dual.
This issue is important because one of the predictions of the correspondence is
the equivalence between the spectrum of string theory on
$\axs$ and the spectrum of anomalous dimensions of gauge invariant
operators. 
There has been a good deal of success recently in comparing the energies  of
semiclassical strings and the anomalous dimensions of the gauge theory operators. 
However, on one side string theory on $AdS_{5}\times S^{5}$ is highly
non-linear and on another the tools at hand to study gauge theory are limited mainly to
perturbation theory. 
Having in mind the duality, i.e. complementarity of the
validity regions of the perturbation theories, the proof of the conjecture even in this
quite well understood case still remains a real challenge. Fortunately, there exists a 
wide region where one can trust quasi-classical considerations
and compare the corresponding spectra on both sides. Integrability provides powerful tools for
calculating the key features on each side of the correspondence.

There are several approaches using simplifications of the non-linear system we are dealing with. 
An important proposal by Berenstein, Maldacena and Nastase \cite{bmn}
made the first step beyond the supergravity approximation in the AdS/CFT 
correspondence. They showed how certain operators in SYM theory 
can be related to string theory in the pp-wave limit in a certain way. 
Namely, it was suggested that string theory in pp-wave background 
gives a good approximation of a certain class of "nearly" chiral operators
and vice versa. Since string theory in the pp-wave 
background is exactly solvable, a lot of information for the AdS/CFT correspondence was 
extracted\footnote{The investigations in pp-wave backgrounds were initiated in \cite{Blau:2002dy}.
 For extensive review see for instance \cite{JPlefka:2004}
\cite{DSadri:2004}.}.

Superstring theory on $\axs$ is integrable at the classical level and there exist strong indications 
 that it happens also at the quantum  level.  Nevertheless, it is a highly  non-trivial task to 
find the exact S-matrix of the theory and to study the duality in its full extent. 
The difficulties are not only of technical nature but also conceptual. Moreover, one must 
keep considerations such that one can check the results on both sides of the correspondence. 
The latter assumes certain limits \cite{GKP},
allowing reliability of the semiclassical results on both sides. In this case the techniques of 
integrable systems have become useful to study the AdS/CFT correspondence in detail. 
For instance, the remarkable observation by Minahan and Zarembo
 \cite{Minahan:2002ve} made it possible to relate the gauge theory side of the correspondence to some 
integrable spin chain 
(at least to leading orders). Thus, the question of relating the string theory side to some
spin chain and then comparing to the gauge theory side becomes very important. 
If so, the integrability of spin chains would provide a powerful tool for investigation
of the AdS/CFT correspondence. 
Discussing the integrability properties of the superstring on $\axs$, one should point out 
several key developments leading to the current understanding. One of them is
the gauge theory S-matrix introduced in \cite{staud:0412} and derived in its full form
in \cite{beis:0511}. In \cite{gleb:04Sm} the authors
consider the Bethe ansatz for quantum strings and introduce the dressing phase as well as the
"symplectic" form of the charges appearing in the exponential\footnote{See \cite{gleb:04Sm}
for further details.}. The power of this
approach was further utilized in \cite{gleb:06:Sm,gleb:06ZF,BHL,janik,BES} leading to important
results.

An important step was done by Maldacena and Hofman \cite{Hofman:2006xt} who were able to map spin chain
"magnon" states to a specific class of rotating semiclassical strings 
on $R\times S^2$ \cite{Kruczenski:2004wg}. This result was soon generalized to magnon bound
states \cite{Dorey1,Dorey2,AFZ,MTT}, dual
to strings on $R\times S^3$ with two  and three non-vanishing angular
momenta. Magnon solutions and the dispersion relations for strings in beta-deformed backgrounds 
were obtained in \cite{Chu:2006ae, Bobev:2006fg}. The applicability of the considerations is restricted
to the case of large quantum numbers, i.e. quasiclassical approximations. Although we assume infinitely large
energy $E$ and momentum $J$, the difference $E-J$ is finite like in the case of the pp-wave limit.
Remarkably enough, due to the integrability on the gauge theory side
one is able to obtain the corresponding S-matrix explicitly \cite{BHL,BES}. 

While similar in dealing with infinitely large quantum numbers, the two pictures, 
namely the pp-wave limit and the giant magnon sector, are a bit different. To see this one can use simple 
scaling arguments. While in the first case the angular momentum $J$ goes to infinity as $\sqrt{\l}\rightarrow\infty$ 
and the product $p\cdotp\l$ is kept fixed (for which $p$ has to be vanishing), in the magnon case
the momentum $p$ is kept fixed. In the first case we have point-like strings and we are at the bottom of the 
$E-J$ scale, while in the second case the string spike is essentially not a point-like object with higher
$E-J$. Following this logic it is natural to look for string solutions interpolating between these two sectors.
In a recent paper Maldacena and Swanson \cite{MS} suggested another limit, called near flat space (NFS) limit, 
which is interpolating between
the two regions above\footnote{One should note that the limit $p\lambda^{1/4}=$const. was 
considered before in \cite{gleb:04Sm} in the context of \cite{GKP}, while the general discussion of this limit
on which our study is based appeared in \cite{MS}.}. In doing a Penrose limit of the geometry we perform a 
certain boost in the vicinity of 
a given null geodesics. The limit is taken in such a way that the product $p\cdotp\sqrt{\l}$ is fixed but the
space becomes of pp-wave type. The authors of \cite{MS} suggest a weaker boosting and the limit is
taken with $p^2\sqrt{\l}$ fixed. The resulting sigma model is much simpler than that of the $\axs$ superstring 
and more complicated than in the pp-wave limit. A remarkable feature of the reduced model is that,
as compared to the pp-wave case, it keeps much more information of the original theory. For instance,
it is integrable with the Lax connection obtained from the original one by taking the limit, and
it possesses the same supersymmetry algebra with a phase originating from the Hopf algebra structure
of the problem. The corresponding S matrix was also conjectured.
The worldsheet scattering of the reduced model and some other properties were
subsequently studied in \cite{Zarembo:2007nfs1,Zarembo:2007nfs2,pul}, see also \cite{Kluson:2007nfs}. 

The discussion above was focused on superstrings on $\axs$ background and its reductions
via certain limits and identifying the corresponding sectors of its dual $\N=4$ SYM theory. 
It is of importance, however, to extend these techniques
to other less symmetric backgrounds that have gauge theory duals. One step in this direction was given
in \cite{BT} where the authors considered the NFS limit of spaces where the $S^5$ part of the geometry is replaced
by certain Sasaki-Einstein spaces. For instance, it is known that spaces like $Y^{pq}$ and $L^{p,q,r}$
have $\N=1$ CFT duals. Remarkably, the near flat space limit of all these spaces is of the same type as in
\cite{MS}. Since the latter is integrable, one can hope that the corresponding theory may contain, 
at least, an integrable subsector. Another step was done in \cite{Kreuzer:2007az} where the near flat space limit
was studied in the Maldacena-Nunez background \cite{MN}\footnote{This geometry was originally obtained in the context of
non-abelian BPS monopoles in gauged supergravity and uplifted to ten dimesions in \cite{Cham-Volk}.}.

Recently, motivated by the possible description of the worldvolume
dynamics of coincident membranes in M-theory in the context of the AdS/CFT correspondence, a new class of
conformally invariant, maximally supersymmetric field theories in 2+1
dimensions has been found \cite{Schwarz:2004yj,ABJM}.
The main feature of these theories is that they contain gauge fields with Chern-Simons-like  kinetic
terms. This development motivated Aharony, Bergman, Jafferis and
Maldacena to propose a new gauge/string duality between an $\mathcal N=6$
 super-conformal Chern-Simons theory (ABJM theory) coupled with
bi-fundamental matter (describing $N$ membranes on $S^7/\mathbb Z_k$) and string
theory on $\cp$. The ABJM model \cite{ABJM} is believed to be holographically dual to M-theory on $AdS_4\times
S^7/\mathbb Z_k$.

The ABJM theory actually consists of two Chern-Simons theories of level $k$
and $-k$, respectively, and each with gauge group $SU(N)$. The two pairs of
chiral superfields transform in the bi-fundamental representation
of $SU(N) \times SU(N)$ and the R-symmetry is $SU(4)$ as it should be for
 $\mathcal N=6$ supersymmetry. It was observed in
\cite{ABJM} that there exists a natural definition of a 't Hooft coupling,
namely, $\lambda = N/k$. 
It was observed that in the 't Hooft limit $N\rightarrow \infty$ with $\lambda$ held fixed,
 one has a continuous coupling  $\lambda$ and the ABJM theory is weakly coupled for $\lambda \ll 1$. 
The ABJM theory is conjectured to be dual to M-theory on $AdS_4\times S^7 / \mathbb Z_k$ with
$N$ units of four-form flux. In the scaling limit $N, k \rightarrow\infty$
with $k \ll N \ll k^5$ the theory can be compactified to type IIA string theory on $AdS_4 \times \mathbb P^3$.
Thus, the AdS/CFT correspondence, which has led to many exciting developments in the duality between type IIB string
theory on $AdS_5\times S^5$ and ${\cal N}=4$ super Yang-Mills theory, is now being extended to $AdS_4/CFT_3$
and is expected to constitute a new example of exact gauge/string theory duality.


The semi-classical string has played an important role in studying various aspects of the
$AdS_5/SYM_4$ correspondence \cite{Bena:2003wd}-\cite{Lee:2008sk}. The developments and successes in this
particular case suggest the methods and tools that should be used to investigate the new emergent duality. 
An important role in these studies plays the integrability. Superstrings on $AdS_4\times\cp$ as a coset were first
studied in \cite{Arutyunov:2008if}\footnote{See also \cite{Stef}} which opens the door for investigation of 
 integrable structures in the theory. Various properties on the gauge theory side and tests on the string theory side,
like rigid rotating strings, pp-wave limit, relations to spin chains,
as well as a pure spinor formulation,  have been considered \cite{Arutyunov:2008if}-\cite{D'Auria:2008cw}.
In these intesive studies many properties were uncovered and impressive results obtained,
but still the understanding of this duality is far from complete.

Inspired by the success of the study of the near flat space limit in the known cases mentioned above, 
in this paper we focus on the ABJM theory. The paper is organized as follows. First we give a brief review of the
near-flat space limit of $\axs$. In the next section we give a concise definition of the Aharony-Bergman-Jafferis-Maldacena 
(ABJM) theory and show how one can make the reduction from membranes to strings on $AdS_4\times\cp$. Next we derive the near-flat 
space limit of the string sigma model on $AdS_4\times\cp$. We conclude with some comments on the results and future directions.

\sect{The Near flat space limit of $\axs$}

In this short section we will review the procedure of taking the near flat space limit of
type IIB string theory on $\axs$ as suggested in \cite{MS}. As we already discussed in the 
introduction, the idea is to weaken the Penrose limit so that more structures of the original theory
get preserved. The first step is to find the null geodesics with respect to which the
pp-wave limit can be taken. We start with the standard metric of $\axs$ in global coordinates
\eq{
ds^2=R^2\lb[-\cosh^2\rho\, dt^2+d\rho^2+\sinh^2\rho\,d\tilde\Omega_3^2+
d\theta^2+\cos^2\theta\,d\psi^2+\sin^2\theta\,d\Omega_3^2\rb]
}
where each of the unit 3-spheres, we take below $S^3$ from the spherical part, can be parameterized by
\eq{
d\Omega_3^2=d\f_1^2+\cos^2\f_1\,d\f_2^2+\sin^2\f_2\,d\f_3^2.
}
The light-like directions are characterized by covariantly constant null vectors
\eq{
\nabla_\m v_\n=0, \quad v_\m v^\m=0.\notag
}
The $S^5$ part of the null geodesics  is parameterized by $\theta=0$ and $\rho=0$, i.e.
in our case this is the equator of the five-sphere. Now we boost the worldsheet coordinates 
and expand about the geodesics ($\dot\psi=1$). This amounts to a redefinition of the fields as follows,
\al{
& t=\sqrt{g}\sigma^++\frac{\tau}{\sqrt{g}} , \quad \psi=\sqrt{g}\sigma^++\frac{\chi}{\sqrt{g}}, \notag \\
& \rho=\frac{z}{\sqrt{g}}, \quad \theta=\frac{y}{\sqrt{g}},
}
where we switched to light-cone worldsheet coordinates $\sigma^+,\sigma^-$. The parameter $g$
is related to the radius of $S^5$ (and $AdS_5$ as well) via $g=R^2/4\pi$.

The near flat space limit is, as in the Penrose limit, taking $g\rightarrow\infty$. In the Lagrangian
the leading terms are divergent, $g\cdot \p_-(\tau-\chi)$, but since they are total derivatives they can be dropped. 
The relevant (bosonic) part of the Lagrangian thus becomes
\eq{
S=4\lb[-\p_+\tau\,\p_-\tau+\p_+\chi\,\p_-\chi+\p_+\vec z\,\p_-\vec z+\p_+\vec y\,\p_-\vec y
 -\vec y^2\,\p_-\chi-\vec z^2\,\p_-\tau \rb]
}
where we skipped the fermionic part and used the notation $\vec z$ for coordinates in $AdS_5$ and
$\vec y$ for those in the $S^5$ part.
The reduced model has two conserved chiral currents
\al{
& j^\chi_+=\p_+\chi-\frac{\vec y^2}{2} + \text{fermions}, \quad \p_-j^\chi_+=0, \notag \\
&j^\tau_+=\p_+\tau+\frac{\vec z^2}{2}- \text{fermions}, \quad \p_-j^\tau_+=0. 
}
The right-moving conformal invariance is preserved since its generator
\eq{
T_{--}=-(\p_-\tau)^2+(\p_-\chi)^2+(\p_-\vec z)^2+(\p_-\vec y)^2+\text{fermions}
}
is conserved, i.e. $\p_+T_{--}=0$. The left-moving conformal invariance however is broken, since 
$T_{++} \varpropto (j^\chi_+-j^\tau_+)$. One can still impose the Virasoro constraints
requiring
\al{
& j^\chi_++j^\tau_+=\p_+(\tau+\chi)+\frac{z^2-y^2}{2}=0, \notag \\
& T_{--}=0.
}
This can also be considered as a gauge fixing condition.

It is useful now to make a change of variables
\eq{
x^+=\sigma^+, \quad x^-=2(\tau+\chi).
}
For completeness we write down the complete gauge fixed Lagrangian
\al{
\L=& 4\Big\{\p_+\vec z\,\p_-\vec z+\p_+\vec y\,\p_-\vec y -\frac{1}{4}(\vec z^2+\vec y^2)
+(\vec y^2-\vec x^2)[(\p_-\vec z)^2+(\p_-\vec y)^2] \notag \\
& +i\psi_+\p_-\psi_+ + i\psi_-\p_+\psi_- + i\psi_-\Pi\psi_+ + i(\vec y^2-\vec x^2)\psi_-\p_-\psi_- \notag \\
& -\psi_-(\p_-z^j\Gamma^j+\p_-y^{j'}\Gamma^{j'})(z^i\Gamma^i-y^{i'}\Gamma^{i'})\psi_- \notag \\
& +\frac{1}{24}\big[\psi_-\Gamma^{ij}\psi_-\psi_-\Gamma^{ij}\psi_-
\psi_-\Gamma^{i'j'}\psi_-\psi_-\Gamma^{i'j'}\psi_-\big]\Big\}.
}
In the last expression a simple rescaling of $\psi_\pm$ was used. The indices $i$ and $i'$ correspond
to the transverse directions in the anti-de Sitter and the spherical parts, respectively.


\sect{Near-flat space limit of string sigma model on $AdS_4\times\cp$}


\paragraph{ABJM and strings on $AdS_4\times \cp$:}

To find the ABJM theory one starts with analysing the M2-brane dynamics governed by the
11-dimensional supergravity action \cite{ABJM}
\eq{
S=\frac{1}{2\kappa_{11}^2}\int dx^{11}\sqrt{-g}\left(R-\frac{1}{2\cdot
4!}F_{\mu\nu\rho\sigma}F^{\mu\nu\rho\sigma}\right)-\frac{1}{12\kappa_{11}^2}
\int C^{(3)}\wedge F^{(4)}\wedge F^{(4)}, 
\label{abjm-1}
}
 where $\kappa_{11}^2=2^7\pi^8 l_p^9$. Solving for the equations of motion  
\eq{
R^\mu_\nu=\frac{1}{2}\left(\frac{1}{3!}F^{\mu\a\beta\gamma}F_{\nu\a\beta\gamma}
-\frac{1}{3\cdot 4!}\delta^\mu_\nu
F_{\a\beta\rho\sigma}F^{\a\beta\rho\sigma}\right), 
\label{abjm-2}
} 
and 
\eq{
\p_{\sigma}(\sqrt{-g}F^{\sigma\mu\nu\xi})=\frac{1}{2\cdot
(4!)^2}\epsilon^{\mu\nu\xi\a_1\dots\a_8}F_{\a_1\dots\a_4}F_{\a_5\dots\a_8},
\label{abjm-3}
}
one can find the M2-brane solutions whose near horizon limit becomes $AdS_4\times S^7$
\eq{ 
ds^2=\frac{R^2}{4}ds^2_{AdS_4}+R^2 ds^2_{S^7}.
\label{abjm-4}
}
In addition we have $N'$ units of four-form flux 
\eq{
 F^{(4)}=\frac{3R^3}{8}\epsilon_{AdS_4}, \quad R=l_p(2^5 N'\pi^2)^{\frac{1}{6}}.
\label{abjm-5}
}
Now one proceeds with considering the quotient $S^7/\mathbb Z_k$ 
acting as $z_i \rightarrow e^{ i { 2 \pi \over k } } z_i$. It is convenient
first to write the metric on $S^7$ as
\eq{
ds^2_{S^7} =  ( d \varphi' + \omega)^2 + ds^2_{CP^3},
\label{abjm-6}
}
where
\al{
& ds^2_{CP^3} = { \sum_i d z_i d \bar z_i  \over r^2 } -
    { | \sum_i z_i d \bar z_i |^2 \over r^4} ~,\quad
 r^2 \equiv \sum_{i=1}^4 |z_i|^2,\notag \\
& d \varphi' + \omega  \equiv  { i \over 2  r^2 } (  z_i d \bar z_i - \bar z_i d z _i ),\quad
d\omega =  J = { i}   d \left({ z_i \over r}\right)  d \left({ \bar z_i \over r }\right).
\label{abjm-7}
}
and then to perform the $\mathbb Z_k$ quotient identifying
$\vf'=\vf/k$ with $\vf\sim \vf+2\pi$ ($J$ is proportional to the K\"ahler form on $\cp$). 
The resulting metric becomes
\eq{
ds^2_{S^7/{\mathbb Z}_k} = { 1 \over k^2 } ( d \vf+ k \o)^2 + ds^2_{CP^3}.
}
One observes that the first volume factor on the right hand side is devided by a factor of $k^2$ as compared to
the initial one. In order to have consistent quantized flux one must impose $N'=kN$ where $N$ is 
the number of quanta of the flux on the quotient.
One should note that the spectrum of the supergravity fields of the final theory is just the projection of the 
initial $AdS_4\times S^7$ onto the $\mathbb Z_k$ invartiant sector. In this setup there is a natural
definition of the {}'t Hooft coupling $\lambda \equiv N/k$. The decoupling limit should be taken
as $N,k\rightarrow \infty$ while $N/k$ is kept fixed.

One can follow now \cite{ABJM} to make a reduction to type IIA with the following final result
\al{
 ds^2_{string} = & { R^3 \over k} ( { 1 \over 4 } ds^2_{AdS_4} + ds^2_{{\mathbb CP}^3 } ),
 \\
 e^{2 \phi} = & { R^3 \over k^3 } \sim { N^{1/2} \over k^{5/2} }= { 1 \over N^2 } \left( {
 N \over k } \right)^{5/2},
 \\
 F_{4} = & { 3 \over 8 }  {  R^3}  \epsilon_4 ,
 \quad
 F_2 =  k d \omega = k J,
 }
 We end up then with the $AdS_4 \times\cp$ compactification
 of type IIA string theory with $N$ units of $F_4$ flux on $AdS_4$ and $k$ units of 
$F_2$ flux on the ${\mathbb CP}^1 \subset\cp$ 2-cycle.

The radius of curvature in string units is $R^2_{str} = { R^3 \over k}  =  2^{5/2} \pi \sqrt{ \lambda}$.
It is important to note that the type IIA approximation is valid in the regime where $k \ll N \ll k^5$.


To fix the notations, we write down the explicit form of the metric on $AdS_4\times\mathbb{CP}^3$
in spherical coordinates. The metric on $AdS_4\times\mathbb{CP}^3$ can be written as \cite{PopeWarner}
\begin{multline}
 ds^2=R^2\left\lbrace \dfrac{1}{4}\left[ -\cosh^2\rho\,dt^2+d\rho^2+\sinh^2\rho\,d\Omega_2^2\right] \right.\\
\left.+d\mu^2+\sin^2\mu\left[d\alpha^2+\dfrac{1}{4}\sin^2\alpha(\sigma_1^2
+\sigma_2^2+\cos^2\alpha\sigma_3^2)+\dfrac{1}{4}\cos^2\mu(d\chi+\sin^2\mu\sigma_3)^2\right]\right\rbrace.
\label{metric}
\end{multline}
Here $R$ is the radius of the $AdS_4$, and $\sigma_{1,2,3}$ are left-invariant 1-forms on an $S^3$, parameterizet by $(\theta,\phi,\psi)$,
\begin{align}
&\sigma_1=\cos\psi\,d\theta+\sin\psi\sin\theta\,d\phi,\notag\\
&\sigma_2=\sin\psi\,d\theta-\cos\psi\sin\theta\,d\phi,\label{S3}\\
&\sigma_3=d\psi+\cos\theta\,d\phi.\notag
\end{align}
The range of the coordinates is
$$0\leq\mu,\,\alpha\leq\dfrac{\pi}{2},\,\,0\leq\theta\leq\pi,\,\,0\leq\phi\leq2\pi,\,\,0\leq\chi,\,\psi\leq4\pi.$$


\paragraph{Taking near-flat space limit of $AdS_4\times\cp$:}

There are several ways to parametrize the $\cp$ part of the spacetime.  One of them is presented in
\eqref{metric} and makes use of the embedding $SU(2)\subset SU(3)\subset SU(4)$. Another way, which puts
an emphasis on an eventual limit to the $S^2\times S^2$ sigma model, is \cite{Nishioka:2008gz} 
\begin{eqnarray}
ds_{CP^{3}}^{2} & = & d\xi^{2}+\cos^{2}\xi\sin^{2}\xi(d\psi+\frac{\cos\theta_{1}}{2}d\varphi_{1}-
\frac{\cos\theta_{2}}{2}d\varphi_{2})^{2}\\
&& +\frac{1}{4}\cos^{2}\xi(d\theta_{1}^{2}
+\sin^{2}\theta_{1}d\varphi_{1}^{2})+\frac{1}{4}\sin^{2}\xi(d\theta_{2}^{2}+\sin^{2}\theta_{2}d\varphi_{2}^{2}).
\end{eqnarray}
The anti-de-Sitter part is described in a standard way
\eq{
ds_{AdS_{4}}^{2}  =  -\cosh^{2}\rho dt²+d\rho^{2}+\sinh^{2}\rho d\Omega_{2}^{2}.
}

The aim of this Section is to study the near-flat space limit of the spacetime. While for the $AdS_4$ par
the procedure is well known from the $AdS_5\times S^5$ case, we will focus mainly on the details for the $\cp$ part.
To this end we introduce the coordinate $\widetilde\psi$ by
 \begin{eqnarray}
\widetilde{\psi} & = & \psi+\frac{\varphi_{1}-\varphi_{2}}{2},\label{redef-psi}
\end{eqnarray}
and make a rescaling of the other coordinates as follows 
\begin{eqnarray}
\rho=\frac{r}{R}, & \theta_{i}=\frac{\sqrt{2}y_{i}}{R}, & \xi=\frac{\pi}{4}+\frac{y_{3}}{2R},
 \quad i=1,2.
\label{eq:scalings}
\end{eqnarray}
Here $\widetilde{\psi}$ is the direction of the null geodesic obtained by taking 
$\rho=\theta_1=\theta_2=0$.
 
Taking the near-flat space limit means taking $R\rightarrow\infty$,
hence we obtain for the metric the following expansion,
\begin{eqnarray}
4ds_{CP^{3}}^{2} & = & \frac{dy_{3}^{2}}{R^{2}}+(1-\frac{y_{3}^{2}}{R^{2}})(d\widetilde{\psi}^{2}-
\frac{y_{1}^{2}}{R^{2}}d\widetilde{\psi}d\varphi_{1}+\frac{y_{2}^{2}}{R^{2}}
d\widetilde{\psi}d\varphi_{2}) \nonumber \\
&& +(\frac{dy_{1}^{2}}{R^{2}}+
\frac{y_{1}^{2}}{R^{2}}d\varphi_{1}^{2})+(\frac{dy_{2}^{2}}{R^{2}} +\frac{y_{2}^{2}}{R^{2}}d\varphi_{2}^{2})+\mathcal{O}(R^{-4})\nonumber \\
ds_{AdS_{4}}^{2} & = & -(1-\frac{r^{2}}{R^{2}})dt^{2}+
\frac{dr^{2}}{R^{2}}+\frac{r^{2}}{R^{2}}d\Omega_{2}^{2}+\mathcal{O}(R^{-4}).
\end{eqnarray}

To take the limit we have to restore the dependence of the metric on the radius $R$, 
which in the string frame metric is
\begin{eqnarray}
ds_{IIA}^{2} & = & R^{2}(ds_{AdS_{4}}^{2}+4ds_{CP^{3}}^{2}).
\end{eqnarray}
In addition it is useful to diagonalize the metric and to make a redefinition of the angles as follows
\begin{align}
\varphi_{1}=\phi_{1}+\frac{\widetilde{\psi}}{2},\quad & \varphi_{2}=\phi_{2}-\frac{\widetilde{\psi}}{2}.
\label{redef-diag}
\end{align}
After all these we end up with
\begin{eqnarray}
ds_{IIA}^{2} & = & R^{2}\{-(1-\frac{r^{2}}{R^{2}})dt^{2}+\frac{dr^{2}}{R^{2}}+\frac{r^{2}}{R^{2}}
d\Omega_{2}^{2}+\frac{dy_{3}^{2}}{R^{2}}+(1-\frac{y_{3}^{2}}{R^{2}})d\widetilde{\psi}^{2}-
\frac{y_{1}^{2}}{4R^{2}}d\widetilde{\psi}^{2}-\frac{y_{2}^{2}}{4R^{2}}d\widetilde{\psi}^{2}\nonumber \\
&& +(\frac{dy_{1}^{2}}{R^{2}}+\frac{y_{1}^{2}}{R^{2}}d\phi_{1}^{2})+
 +(\frac{dy_{2}^{2}}{R^{2}}+\frac{y_{2}^{2}}{R^{2}}d\phi_{2}^{2})+\mathcal{O}(R^{-3})\}.
\end{eqnarray}

As it was argued in \cite{MS,BT,Kreuzer:2007az}, in order to describe the region connecting the
giant magnon sector and the pp-wave limit of the string sigma model one must use a specific ansatz for
the variables describing the null geodesic.
We use the following ansatz for $t$ and  $\widetilde{\psi}$
\begin{align}
t=k_{t}R\sigma^{+}+\frac{\tau}{R},\qquad & \widetilde{\psi}=kR\sigma^{+}+\frac{\chi}{R}.
\label{eq:geodesic}
\end{align}
Substituting into the Lagrangian we find
 $\mathcal{L}=\partial_{+}X^{\mu}\partial_{-}X^{\nu}G_{\mu\nu}$
\begin{eqnarray*}
\mathcal{L} & = & -r^{2}k_{t}\partial_{-}\tau-\partial_{+}\tau\partial_{-}\tau+
\partial_{+}r\partial_{-}r+r^{2}(\partial_{+}\theta_{3}
\partial_{-}\theta_{3}+\sin\theta_{3}^{2}\partial_{+}\varphi_{3}
\partial_{-}\varphi_{3}) \\
&& +\sum_{i=1,3}\partial_{+}y_{i}
\partial_{-}y_{i}+y_{1}^{2}\partial_{+}\phi_{1}\partial_{-}\phi_{1}+\\
 &  & +\partial_{+}\chi\partial_{-}\chi-(\frac{y_{2}^{2}}{4}+
\frac{y_{1}^{2}}{4}+y_{3}^{2})k\partial_{-}\chi+y_{2}^{2}\partial_{+}\phi_{2}\partial_{-}\phi_{2}.
\end{eqnarray*}

The Virasoro constraints fix $k=k_{t}$, which, by making use of the residual conformal symmetry, 
is set to one. After obvious redefinitions of coordinates we obtain for the reduced Lagrangian
\begin{eqnarray}
\mathcal{L} & = & -r^{2}\partial_{-}\tau-\partial_{+}\tau\partial_{-}\tau+\partial_{+}\vec{r}
\partial_{-}\vec{r}+\partial_{+}y_{3}\partial_{-}y_{3}+\partial_{+}
\vec{y}_{1}\partial_{-}\vec{y}_{1} \nonumber \\
& & +\partial_{+}\vec{y}_{2}\partial_{-}
\vec{y}_{2}+\partial_{+}\chi\partial_{-}\chi 
 -(\frac{\vec y_{2}^{2}}{4}+\frac{\vec y_{1}^{2}}{4}+y_{3}^{2})\partial_{-}\chi.
\nonumber 
\end{eqnarray}

The equations of motion following from the reduced Lagrangian are
\begin{eqnarray}
0 & = & \partial_{+}\partial_{-}\tau+\frac{1}{2}\partial_{-}r^{2},\nonumber \\
0 & = & \partial_{+}\partial_{-}\chi-(\frac{\partial_{-}y_{2}^{2}}{8}+
\frac{\partial_{-}y_{1}^{2}}{8}+\frac{\partial_{-}y_{3}^{2}}{2}),\nonumber \\
0 & = & \partial_{+}\partial_{-}r_{i}+r_{i}\partial_{-}\tau,\qquad i=1,2,3\\
0 & = & \partial_{+}\partial_{-}y_{a_{j}}+\frac{y_{a_{j}}}{4}\partial_{-}\chi,\quad a,j=1,2\nonumber \\
0 & = & \partial_{+}\partial_{-}y_{3}+y_{3}\partial_{-}\chi.\nonumber 
\end{eqnarray}
As in the case studied in \cite{MS,BT,Kreuzer:2007az}, one
 obtains the following chiral conserved currents
 \begin{eqnarray}
j_{+}^{\tau} & = & \partial_{+}\tau+\frac{1}{2}r^{2},\\
j_{+}^{\chi} & = & \partial_{+}\chi-(\frac{\vec y_{2}^{2}}{8}+\frac{\vec y_{1}^{2}}{8}+\frac{y_{3}^{2}}{2}).
\end{eqnarray}
Now let us examine the Virasoro constraints.
We start with the energy momentum tensor of the original theory
 \begin{eqnarray}
T_{++} & = & G_{\mu\nu}\partial_{+}X^{\mu}
\partial_{+}X^{\nu},\qquad T_{--}=G_{\mu\nu}\partial_{-}X^{\mu}\partial_{-}X^{\nu}.
\end{eqnarray}

We use the rescalings (\ref{eq:scalings},\ref{eq:geodesic}) and definitions
(\ref{redef-psi}) and (\ref{redef-diag}) 
 to obtain the corresponding expression for the reduced model
 \begin{eqnarray}
T_{++} & \rightarrow & -2\partial_{+}\tau-r^{2}+2\partial_{+}
\chi-(\frac{y_{2}^{2}}{4}+\frac{y_{1}^{2}}{4}+y_{3}^{2}),\\
T_{--} & \rightarrow & \frac{1}{R^{2}}(-\partial_{-}\tau\partial_{-}\tau+
\partial_{-}\chi\partial_{-}\chi+\partial_{-}\vec{r}\partial_{-}\vec{r} \\
& & +\partial_{-}y_{3}\partial_{-}y_{3}+\partial_{-}\vec{y}_{1}\partial_{-}\vec{y}_{1}+
\partial_{-}\vec{y}_{2}\partial_{-}\vec{y}_{2}).
\end{eqnarray}

We observe that  $T_{++}\varpropto j_{+}^{\chi}-j_{+}^{\tau}$.
 The meaning of this observation is that the left-moving conformal symmetry is broken and is 
replaced by chiral symmetries which are generated by $j_{+}^{\tau}$ and $j_{+}^{\chi}$.
 We note that at the same time the right-moving conformal symmetry remains unbroken.
The right conformal symmetry is generated by the Virasoro generator having in our case the form
\eq{
T_{--} := -\partial_{-}\tau\partial_{-}\tau+\partial_{-}\chi\partial_{-}\chi+\partial_{-}\vec{r}
\partial_{-}\vec{r}+\partial_{-}y_{3}\partial_{-}y_{3}
+\partial_{-}\vec{y}_{1}\partial_{-}\vec{y}_{1}+\partial_{-}\vec{y}_{2}\partial_{-}\vec{y}_{2}.
}

The next step is to obtain the gauge fixed action. This can be done in a standard way, namely,
to fix the gauge, we impose the following conditions
\begin{eqnarray}
T_{--} & = & 0,\\
j_{+}^{\tau}+j_{+}^{\chi} & = & \partial_{+}\tau+\partial_{+}\chi+\frac{1}{2}r^{2}-
(\frac{\vec y_{2}^{2}}{8}+\frac{\vec y_{1}^{2}}{8}+\frac{y_{3}^{2}}{2})=0.
\end{eqnarray}

To this end we choose coordinates
\begin{equation}
x^{+}\equiv\sigma^{+},\qquad x^{-}\equiv2(\tau+\chi),
\end{equation}
so that the derivatives take the form
 \begin{equation}
\partial_{\sigma^{+}}=\partial_{x^{+}}+[-r^{2}+(\frac{\vec y_{2}^{2}}{4}+
\frac{\vec y_{1}^{2}}{4}+y_{3}^{2})]\partial_{x^{-}},\quad\partial_{\sigma^{-}}=
2[\partial_{\sigma^{-}}(\tau+\chi)]\partial_{x^{-}}.
\end{equation}

With this redefinitions we can rewrite the EOM for dynamical variables
 $\vec{r},\vec{y_{1}},\vec{y_{2}}$ and $y_{3}$. The latter take the form
\begin{eqnarray}
0 & = & \partial_{x^{-}}(\partial_{x^{+}}+[-r^{2}+(\frac{y_{2}^{2}}{4}+\frac{y_{1}^{2}}{4}+y_{3}^{2})]
\partial_{x^{-}})r_{i}+r_{i}\,(\frac{1}{4}+\partial_{x^{-}}\vec{r}\partial_{x^{-}}\vec{r}\nonumber \\
 &  &  +
\partial_{x^{-}}y_{3}\partial_{x^{-}}y_{3}+\partial_{x^{-}}\vec{y}_{1}
\partial_{x^{-}}\vec{y}_{1}+\partial_{x^{-}}\vec{y}_{2}\partial_{x^{-}}\vec{y}_{2}),
\qquad i=1,2,3\\
0 & = & \partial_{x^{-}}(\partial_{x^{+}}+[-r^{2}+(\frac{y_{2}^{2}}{4}+
\frac{y_{1}^{2}}{4}+y_{3}^{2})]\partial_{x^{-}})y_{a_{j}}+\frac{y_{a_{j}}}{4}(\frac{1}{4}
-\partial_{x^{-}}\vec{r}\partial_{x^{-}}\vec{r} \nonumber \\
& & -\partial_{x^{-}}y_{3}\partial_{x^{-}}y_{3}-\partial_{x^{-}}\vec{y}_{1}
\partial_{x^{-}}\vec{y}_{1}-\partial_{x^{-}}\vec{y}_{2}\partial_{x^{-}}\vec{y}_{2}),\qquad a,j=1,2\\
0 & = & \partial_{x^{-}}(\partial_{x^{+}}+[-r^{2}+(\frac{y_{2}^{2}}{4}+
\frac{y_{1}^{2}}{4}+y_{3}^{2})]\partial_{x^{-}})y_{3}+\nonumber \\
 &  & +y_{3}(\frac{1}{4}-\partial_{x^{-}}\vec{r}\partial_{x^{-}}\vec{r}-
\partial_{x^{-}}y_{3}\partial_{x^{-}}y_{3}-\partial_{x^{-}}\vec{y}_{1}
\partial_{x^{-}}\vec{y}_{1}-\partial_{x^{-}}\vec{y}_{2}\partial_{x^{-}}\vec{y}_{2}),
\end{eqnarray}
where we used the gauge fixing condition
\begin{equation}
T_{--}=0\Longrightarrow\frac{1}{2}\partial_{x^{-}}(\tau-\chi)=\partial_{x^{-}}\vec{r}
\partial_{x^{-}}\vec{r}+\partial_{x^{-}}y_{3}\partial_{x^{-}}y_{3}+\partial_{x^{-}}\vec{y}_{1}
\partial_{x^{-}}\vec{y}_{1}+\partial_{x^{-}}\vec{y}_{2}\partial_{x^{-}}\vec{y}_{2}.
\end{equation}

These equations can be thought of as derived from the following effective
 Lagrangian
\begin{eqnarray}
\mathcal{L}_{eff} & = & \partial_{+}\vec{r}\partial_{-}\vec{r}+\partial_{+}\vec{y}_{2}\partial_{-}\vec{y}_{2}+
\partial_{+}\vec{y}_{1}\partial_{-}\vec{y}_{1}+\partial_{+}y_{3}\partial_{-}y_{3}+\nonumber \\
 &  & +[-r^{2}+(\frac{\vec y_{2}^{2}}{4}+\frac{\vec y_{1}^{2}}{4}+y_{3}^{2})](\partial_{-}\vec{r}\partial_{-}
\vec{r}+\partial_{-}\vec{y}_{2}\partial_{-}\vec{y}_{2}+\partial_{-}\vec{y}_{1}\partial_{-}
\vec{y}_{1}+\partial_{-}y_{3}\partial_{-}y_{3})+\nonumber \\
 &  & -\frac{\vec r^{2}}{4}-\frac{\vec y_{2}^{2}}{16}-\frac{\vec y_{1}^{2}}{16}-\frac{y_{3}^{2}}{4}.
\end{eqnarray}

For this effective Lagrangian we introduce $\sigma^\pm=\tau\pm\sigma$ and
hence we obtain the following momenta ( with  $4\mathcal{L}=\mathcal{L}'$)
\begin{eqnarray*}
p_{r_{i}} & = & \frac{\partial\mathcal{L}'}{\partial(\partial_{\tau}r_{i})}
 =  2\partial_{\tau}r_{i}+[-r^{2}+(\frac{y_{2}^{2}}{4}+
\frac{y_{1}^{2}}{4}+y_{3}^{2})](2\partial_{\tau}r_{i}-2\partial_{\sigma}r_{i}),\\
p_{y_{i}} & = & \frac{\partial\mathcal{L}'}{\partial(\partial_{\tau}y_{i})}
  =  2\partial_{\tau}y_{i}+[-r^{2}+(\frac{y_{2}^{2}}{4}+
\frac{y_{1}^{2}}{4}+y_{3}^{2})](2\partial_{\tau}y_{i}-2\partial_{\sigma}y_{i}),
\end{eqnarray*}
with the canonical Poisson brackets
\eq{
\{r^i(\sigma),p_{rj}(\sigma')\}=\delta_j^i\delta(\sigma-\sigma'), \quad
\{y^i(\sigma),p_{yj}(\sigma')\}=\delta_j^i\delta(\sigma-\sigma').
}
From here we find
\begin{eqnarray*}
 \p_\tau\vec r=\frac{\frac{1}{2}\vec p_{r}+(y_3^2-r^2+\frac{y_1^2+y_2^2}{4})
\p_\sigma\vec r}{1+y_3^2-r^2+\frac{y_1^2+y_2^2}{4}},\\
\p_\tau\vec y=\frac{\frac{1}{2}\vec p_{y}+(y_3^2-r^2+
\frac{y_1^2+y_2^2}{4})\p_\sigma\vec y}{1+y_3^2-r^2+\frac{y_1^2+y_2^2}{4}}.
\end{eqnarray*}

The Hamiltonian then reads 
\ml{
\H=\frac{1}{4}\frac{\vec p_r^2+\vec p_y^2}{1+y_3^2-r^2+\frac{y_1^2+y_2^2}{4}}+
\frac{(y_3^2-r^2+\frac{y_1^2+y_2^2}{4})(\vec p_r.\p_\sigma\vec r+\vec p_y.\p_\sigma\vec y)}
{1+y_3^2-r^2+\frac{y_1^2+y_2^2}{4}}\\
+\frac{(\p_\sigma\vec r)^2+(\p_\sigma\vec y)^2}{1+y_3^2-r^2+\frac{y_1^2+y_2^2}{4}}
+ \frac{r^{2}}{4}+\frac{y_{2}^{2}}{16}+\frac{y_{1}^{2}}{16}+\frac{y_{3}^{2}}{4}.
\label{hamilt-2}
}
or, equivalently
\ml{
\H=\frac{\frac{\vec p_r^2+\vec p_y^2 }{4}+
(y_3^2-r^2+\frac{y_1^2+y_2^2}{4})(\vec p_r.\p_\sigma\vec r+\vec p_y.\p_\sigma\vec y)
+(\p_\sigma\vec r)^2+(\p_\sigma\vec y)^2    }{1+y_3^2-r^2+\frac{y_1^2+y_2^2}{4}} \\
+ \frac{r^{2}}{4}+\frac{y_{2}^{2}}{16}+\frac{y_{1}^{2}}{16}+\frac{y_{3}^{2}}{4}.
\label{hamilt-2}
}

One can use the Hamiltonian to further investigate the properties of the reduced model and
this would be an interesting line to proceed.

\sect{Conclusions}

As we discussed in the Introduction, string backgrounds other than $\axs$ are of great interest,
especially because their gauge theory duals are closely related to QCD or are of particular importance for
the prove of the holographic conjecture. 
Typically the string backgrounds define highly non-linear sigma models difficult to deal with. 
It is then natural to look for limiting procedures providing useful simplification but still allowing to extract
non-trivial information for the models. One of the limits widely used in the last years is the so-called pp-wave limit. 
It is very attractive because the resulting theory is exactly sovable, describes non-trivial properties and 
one be quantized. The issues related to the validity of the correspondence 
are have been  also addressed. Although this limit possesses very nice properties like integrability, some of 
the important properties and important structures of the original theory get lost. 
One possible way out is to weaken the pp-wave limit
and to preserve many of the key features lost on the way. Such a procedure for the superstring theory on 
the $\axs$ background
was proposed by Maldacena and Swanson. They called this procedure near-flat space limit.

In this paper we tested the near-flat space limit for the case of string theory on
$AdS_4\times\cp$. Being the string dual of a very non-trivial gauge theory, 
$\mathcal N=6$ CS theory, superstring theory on this  
background is of particular importace.
As in the $\axs$ case, it is conjectured that
many of the key features of the original theory survive the near-flat space limit 
and can be studied in the reduced model. Our considerations 
show that the reduced model in this background is of the same type as the one obtained from the NFS limit of
$\axs$ \cite{MS}, Sasaki-Einstein \cite{BT} and Maldacena-Nunez \cite{Kreuzer:2007az}. 
The bosonic part of the gauge fixed lagrangian of the reduced model contains again ``mass terms''. 
The latter are composed differently from  the case considered in \cite{MS,BT,Kreuzer:2007az}. 
In the Appendix we obtain the Lax pair for the reduced $\cp$ model which proves the integrability
at the classical level. 
The similarity of the structure gives more arguments towards the conjectured
universality of the near flat space limit of string theory on 
such backgrounds and their integrability. Therefore, the techniques developed in 
\cite{Zarembo:2007nfs1,Zarembo:2007nfs2,pul} will be also universal within the class 
of the NFS reduced models and one can apply them here.

There are several directions for further generalizations of this study. 
First of all, it would be interesting, following the close similarity of the structure
of the Lagrangian obtained here and in \cite{MS}, to
find the explicit form of the Lax connection for the full string gauge fixed model and to perform 
an analysis along the lines of, say \cite{Kluson:2007nfs}. 
Another direction would be to apply
these techniques to the study of string sigma models in so called beta-deformed backgrounds introduced
first in \cite{Lunin:2005jy},
assuming that the NFS limit interpolates between pp-wave and magnon sectors of the theory
\cite{Chu:2006ae,Bobev:2006fg}.
In this paper we considered only the bosonic part of the theory, which is analogous to the $\axs$ case
in many ways. The near-flat space limit of the fermionic part is much more difficult due to the presence of non-trivial
RR fields and it would be very important to extend the consideration to that case.
Although there is some progress in our understanding of the role of the integrable structures
in AdS/CFT, or more generally string/gauge theory duality, much has still to be done.

\bigskip
\leftline{\bf Acknowledgments}
\smallskip
We thank C. Mayrhofer for useful discussion at an early stage of the project.
R.R. acknowledges the warm hospitality and stimulating atmosphere at the ITP of TU Wien and the
Erwin Schr\"{o}dinger International Institute for Mathematical Physics (ESI).
M.S. acknowledges also ESI for the Junior Research Fellowship during which this studies were
initiated.
This work was supported in part by the Austrian Research Fund FWF grant \# P19051-N16.
R.R. is partly supported by the Bulgarian Research Fund NSF grant VU-F-201/06.


\section*{Appendix: Remark on the Lax pair of the reduced $\mathbb{CP}^n$ sigma model}

\def\theequation{A.\arabic{equation}}
\setcounter{equation}{0}
\begin{appendix}

The construction of the Lax pair for the reduced $O(n)$ model was dicussed in \cite{MS}. One way to do that 
in our case is to
take the near-flat space limit of the connection in the $\mathbb{CP}^n$ sigma model. We will pursue this issue elsewhere, 
but in this appendix we will take advantage of what has been done for the $O(n+2)$ case considering 
$\mathbb{CP}^n$ as $\mathbb{CP}^n \simeq S^{n+1}/U(1)$. Here we will restrict the considerations to
the case of $cp$ sigma model as embedded into a seven sphere. We choose the generator $J^{12}$ 
which mixes (1,2) plane with four of the directions, say $J^{\pm i}, \; i=1,\cdots,4$. 
In addiditon, one of the directions, say $y_{5}$\footnote{Here we rename $y_{a_i}$ as 
$y_i$ and $y_3$ becomes $y_5$.}, is associated with a generator $J^{\pm 5}$ 
commuting with all $J^{\pm i}$ exept $J^{12}$.
To construct the Lax connection we will use the following commutation relations:
\al{
& [J^{12},J^{\pm i}]=\pm J^{\pm i}, \quad [J^{12},J^{\pm 5}]=\pm J^{\pm 5} \notag \\
& [J^{+ i},J^{- j}]=-\d_{ij}J^{12}-J^{ij}, \quad [J^{- i},J^{+j}]=\d_{ij}J^{12}-J^{ij}\\
& [J^{+5},J^{- 5}]=-J^{12}, \quad [J^{\pm}, J^{\mp 5}]=0. \notag 
}

Now we can construct the flat connection with spectral parameter $\o$ 
\al{
& d+\mathcal A=d+A_+d\s^++A_-d\s^-, \quad d^2A+A\wedge A=0. \notag \\
 A_+=  &\frac{i}{\sqrt{2}}\lb[-e^{-i\s^+\o}\frac{y^i}{4}J^{+i}+e^{i\s^+\o}
\frac{y^i}{4}J^{-i}-e^{-i4\s^+\o}y^5J^{+5}+e^{i4\s^+\o}y^5J^{-5}\rb] \label{lax-p} \\
  A_-= &\frac{1}{4\o}\lb[-i\p_-\chi\; J^{12}+\frac{1}{\sqrt{2}}e^{-i\s^+\o}\p_-y^iJ^{+i}
+\frac{1}{\sqrt{2}}e^{-i4\s^+\o}\p_-y^5J^{+5} \right. \notag \\
 & +\left. \frac{1}{\sqrt{2}}e^{i\s^+\o}\p_-y^iJ^{-i}
+\frac{1}{\sqrt{2}}e^{i4\s^+\o}\p_-y^5J^{-5}  \rb] 
}
A simple check shows that the equations of motion are equivalent to the flatness of the connection.
An interesting feature of the model is that one can use the new variables introduced for the gauge fixed
effective Lagrangian to obtain its Lax connection. Indeed, having in mind that
\eq{
d\s^+=dx^+, \quad 2\;d\s^-(\frac{1}{4}+\p_{\s^-}\chi)=dx^--(\frac{y_i^2}{4}
+y_5^2)dx^+
}
and using the constraints we find the new connection in terms of the
transverse degrees of freedom\footnote{In the case under consideration ($j_\tau\equiv 0$),
after adding a trivial time direction, the
Virasoro constraints are $j_\chi=0$ and $T_{--}=\frac{1}{16}$.}
\eq{
d+A_+d\s^++A_-d\s^-\quad \rightarrow\quad  d+A_+dx^++\tilde A(dx^--(\frac{y_i^2}{4}
+y_5^2)dx^+),
}
where
\ml{
\tilde A:= \frac{1}{4\o}\lb[-i\Big(\frac{1}{4}-(\p_{x^-}y_5)^2-
(\p_{x^-}y_i)^2\Big)J^{12} 
 +\frac{1}{\sqrt{2}}e^{-i\s^+\o}\p_-y^iJ^{+i}\right.  \\
+\frac{1}{\sqrt{2}}e^{-i4\s^+\o}\p_-y^5J^{+5}
  +\left.\frac{1}{\sqrt{2}}e^{i\s^+\o}\p_-y^iJ^{-i}
+\frac{1}{\sqrt{2}}e^{i4\s^+\o}\p_-y^5J^{-5} \rb].
}
A simple gauge transformation with $g=exp(-i\frac{x^-}{4\o}J^{12})$ removes the constant part
and one ends up with
\eq{
 d+A_+'dx^++\tilde A'\Big(dx^--(\frac{y_i^2}{4}+y_5^2)dx^+\Big),
}
where
\ml{
 A_+'= \frac{i}{\sqrt{2}}\lb[-e^{-ix^+\o +i\frac{x^-}{4\o}}\frac{y^i}{4}J^{+i}
-e^{-i4x^+\o +i\frac{x^-}{4\o}}y^5J^{+5} \right. \\
\left. +e^{ix^+\o -i\frac{x^-}{4\o}}\frac{y^i}{4}J^{-i}+e^{i4x^+\o -i\frac{x^-}{4\o}}y^5J^{-5})\rb],
}
\ml{
\tilde A'= \frac{1}{4\o}\lb[i\Big((\p_{x^-}y_5)^2+
(\p_{x^-}y_i)^2\Big)J^{12} 
 +\frac{1}{\sqrt{2}}e^{-ix^+\o +i\frac{x^-}{4\o}}\p_-y^iJ^{+i}\right. \\
+\frac{1}{\sqrt{2}}e^{-i4x^+\o +i\frac{x^-}{4\o}}\p_-y^5J^{+5} 
  +\frac{1}{\sqrt{2}}e^{ix^+\o -i\frac{x^-}{4\o}}\p_-y^iJ^{-i}\\
\left.+\frac{1}{\sqrt{2}}e^{i4x^+\o -i\frac{x^-}{4\o}}\p_-y^5J^{-5} \rb].
}
This form of the Lax connection is not unique and maybe not be the most useful one, 
but we will further study this issue  in another paper.

\end{appendix}


\end{document}